\documentclass{svproc}
\usepackage{url}
\usepackage{graphicx}
\usepackage{caption}
\usepackage{subcaption}
\usepackage{rotating}
\usepackage{array}

\usepackage{amsfonts}

\begin{document}
\mainmatter              
\title{The Structure of Interdisciplinary Science: Uncovering and Explaining Roles in Citation Graphs}
\titlerunning{Roles in Citation Graphs}  
%
\author{Eoghan Cunningham \inst{1,2} \and Derek Greene\inst{1,2}}
\authorrunning{Cunningham and Greene.} 
%
\tocauthor{Eoghan Cunningham, Derek Greene}
\institute{School of Computer Science, University College Dublin, Ireland
\and Insight Centre for Data Analytics, University College Dublin, Ireland}

\maketitle              

\begin{abstract}
Role discovery is the task of dividing the set of nodes on a graph into classes of structurally similar roles. Modern strategies for role discovery typically rely on graph embedding techniques, which are capable of recognising complex local structures. However, when working with large, real-world networks, it is difficult to interpret or validate a set of roles identified according to these methods. In this work, motivated by advancements in the field of explainable artificial intelligence (XAI), we propose a new framework for interpreting role assignments on large graphs using small subgraph structures known as graphlets. We demonstrate our methods on a large, multidisciplinary citation network, where we successfully identify a number of important citation patterns which reflect interdisciplinary research. 
\end{abstract}
\section{Introduction}

In light of the perceived importance of interdisciplinary research, many studies have been conducted that quantify article interdisciplinarity in an effort to identify relevant research trends and to explore their impact. The most widely-accepted methods for measuring interdisciplinarity assess \emph{knowledge integration} or \emph{knowledge diffusion} using citation information, thus measuring interdisciplinarity as some function of the balance, diversity, and dissimilarity of the disciplines identified in an article's cited papers \cite{porter2009science,rafols2010diversity} or citing papers \cite{porter1985indicator,van2015interdisciplinary}.
To implement these metrics, each paper in a research corpus must be assigned to an explicit research topic or subject category, for which sources are numerous, inconsistent, and sometimes unavailable. Subject categories are most commonly assigned to papers according to the journals in which they are published. However, these assignments rarely agree with underlying citation community structure \cite{porter2009science}. 
There is evidence that interdisciplinary research can be identified in a corpus according solely to the citation structure. Specifically, it has been shown that frameworks that encode the \emph{structural role} of articles in a citation graph can predict interdisciplinary interactions more accurately than those that encode only the proximity between papers \cite{cunningham2022assessing}. In light of this, we explore the potential for modern graph learning methods to identify the citation structures associated with interdisciplinary research.

Numerous approaches have been proposed in the literature for the task of structural role discovery on graphs, where nodes on a graph are divided into classes of \emph{structurally equivalent} nodes \cite{rossi2014role}. Early approaches in this area relied on graph sub-structures known as \emph{graphlets} or \emph{motifs} \cite{milo2002network}. In large graphs, where we wish to identify higher-order structural features, counting large graphlets is very expensive. The majority of recent approaches employ a representation learning technique known as node embedding, which substitutes graphlets with alternative structural features like degree distributions \cite{ribeiro2017struc2vec} and diffusion wavelet patterns \cite{donnat2018learning}. These methods are designed to learn higher-order structural relationships than those that can be discovered by small graphlets. However, in many cases, these alternatives approaches come at the cost of interpretability. When applied to graphs that are too large to be visualised reasonably, it is difficult to understand the substantive meaning of a given set of structural roles. 

While embedding methods for role analysis have previously been shown to be capable of grouping nodes into known roles or structures (such as those in synthetic graphs or transport networks \cite{donnat2018learning,ribeiro2017struc2vec}), it remains unclear as to how these roles should be interpreted or validated when applied to real-world graphs with unknown roles. Moreover, different role discovery methods learn different sets of structural roles, depending on the application many or none of these clusterings may be valid. As such, it is critical that we can compare the roles discovered by different methods.

The core contribution of this work is a new framework for explaining a set of discovered roles  using graphlet \emph{orbits}, as described in detail in Section \ref{sec:methods}. Later in Section \ref{sec:application} we apply this framework to a large, multidisciplinary citation network to extract and interpret sets of structural roles. In the case of each paper, we compute Rao-Stirling diversity scores to indicate an articles interdisciplinarity. Crucially, in addition to interpreting a set of candidate roles, we explore the distributions of IDR scores assigned to papers in different clustering to assess if any of the candidate clusterings have grouped papers according to their interdisciplinarity.

\section{Related Work}
\subsection{Measuring Interdisciplinarity}

\emph{Interdisciplinary research} is most commonly defined as research activity which integrates expertise, data or methodologies from two or more distinct disciplines. Accordingly, many studies assign an interdisciplinary research (IDR) score to a research article, which is calculated as a function of the balance, diversity and dissimilarity of the disciplines identified in the articles reference list \cite{porter2009science,rafols2010diversity}. Alternatively, some studies compute a similar score on the disciplines identified in an article's \emph{citing} literature, instead measuring IDR according to an articles impact/influence across multiple disciplines \cite{porter1985indicator,van2015interdisciplinary}. A popular function for measuring IDR is the Rao-Stirling Diversity index \cite{stirling2007general}
\begin{equation}
\label{eq:r_s_diversity}
    D = \sum_{i,j (i\neq j)}p_ip_jd_{ij}
\end{equation}
where IDR is measured as a pairwise summation over all pairs of disciplines identified in a set of articles cited by (or citing) some focal paper. Here $p_i$ and $p_j$ represent the proportion of references to disciplines $i$ and $j$ respectively, while $d_{ij}$ refers to some precomputed distance between the disciplines. This metric, and its many variants, are reliant on explicit topic or discipline categories for research papers, such as those provided by Web of Science, Scopus, or Microsoft Academic. Such explicit categorisations for research papers, especially those assigned at a journal level, are problematic \cite{abramo2018comparison,milojevic20science}. Moreover, the inconsistencies evident across many of these subject taxonomies \cite{shen2019node2vec} may confirm that no singular, correct categorisation exists.  However, recent graph learning methods may be capable of identifying the citation structures associated with interdisciplinary research, without any knowledge of explicit discipline categorisation.

\subsection{Local Structure and Role Embeddings}

\begin{figure}[!b]
    \centering
\includegraphics[width=\linewidth]{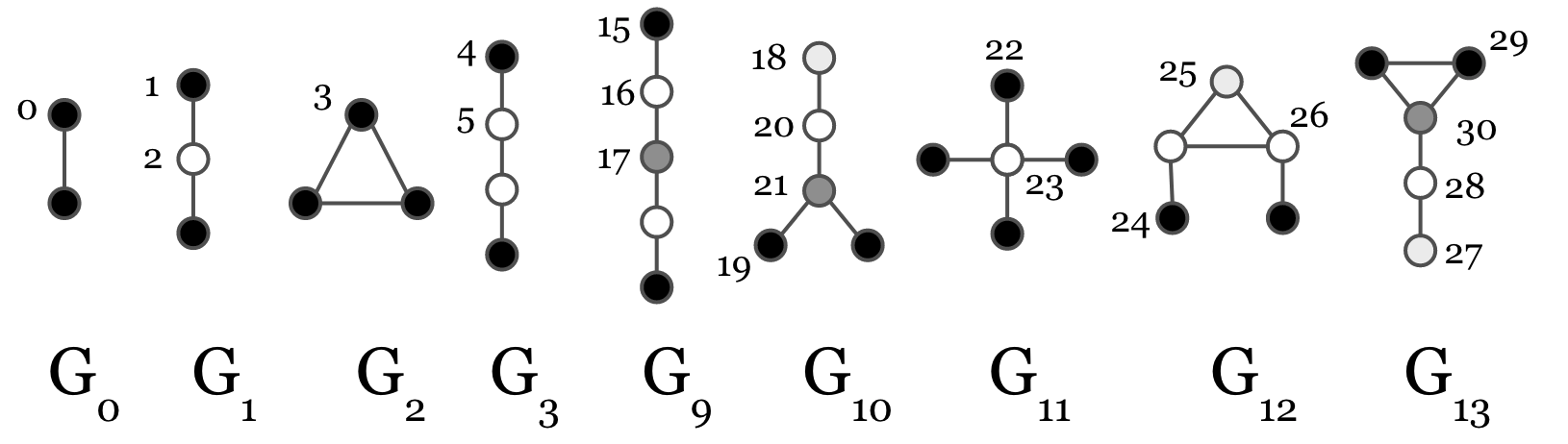}
    \caption{Graphlets with 2, 3, 4, and 5 nodes. Graphlets and orbits are enumerated according to \cite{prvzulj2007biological}}
    \label{fig:graphlets}
\end{figure}

Research developed in social science studied local graph structures using small graph patterns such as triads, cycles and stars \cite{moreno1934shall}. More recent research in computational science further developed these methods and proposed the term `motif' -- a subgraph pattern (or \emph{graphlet}) which is significantly over-represented in a graph \cite{milo2002network}. Motifs, and graphlets represent a powerful tool for expressing graph structure, and have been employed in graph learning tasks such as node classification and anomaly detection \cite{cunningham2013motifs}. Figure \ref{fig:graphlets} illustrates a subset of graphlets with 2, 3, 4 and 5 nodes, and includes each of the distinct orbits on these graphlets, as they were enumerated by \cite{prvzulj2007biological}.

\emph{Role discovery} is the task of grouping nodes which share similar structural patterns in a graph into distinct classes \cite{rossi2014role}. Many modern approaches to role discovery rely on graph embedding, where nodes are transformed into low-dimensional vector representations \cite{ahmed2019role2vec,henderson2012rolx,ribeiro2017struc2vec,donnat2018learning}. Specifically, graph embedding methods for the purpose of role discovery (often called ``role embedding'') learn dense vector representations for nodes such that nodes that are structurally similar will be embedded nearby in the embedding space (i.e., will have similar representations). A clustering of the role embedding space thus represents a set of discovered roles. However, if this network cannot be visualised, it is difficult to interpret the roles. Moreover, with numerous approaches to role embedding (e.g. \cite{ahmed2019role2vec,donnat2018learning,henderson2012rolx,ribeiro2017struc2vec}), and many possible clusterings of each embedding space, we require some approach to explain a set of discovered roles so that they can be compared and validated.

\subsection{Explanation via Surrogate Models} 

In recent years, much research has been conducted in the field of \emph{explainable artificial intelligence} (XAI) towards a goal of understanding and explaining so-called `\emph{black box}' models \cite{adadi2018peeking}. One popular approach is to use a global surrogate model: ``an interpretable model that is trained to approximate the predictions of a black box model'' \cite{molnar2020interpretable}. As such, the decisions made by a previously uninterpretable system can be explained through interpretations coming from the surrogate model. 

Some classification models (such a logistic regression models or decision tree-based classifiers) are interpretable by definition, as any input feature's effect on the models classification can be measured (for example, using regression coefficients). However, many model-agnostic methods of interpretation have also been developed. Generally, these methods propose different means of perturbing the input data, and exploring its effect on the models output. For example, \emph{Partial Dependence Plots} (PDPs) \cite{friedman2001greedy} offer a graphical representation of a models prediction as a function of its arguments or inputs. Broadly speaking, a PDP shows how a feature influences the prediction of a model, on average, for every value that it can take. The \emph{Accumulated Local Effects} plot (ALE) \cite{apley2020visualizing} provides a similar illustration of a features effect, except it accounts for possible correlations between features, and as a result is more robust than the PDP in some applications. In 2021, global surrogate models were employed to provide feature-based explanations for communities detected on graphs \cite{sadler2021selecting}. Specifically, this work assessed the importance of different graph features using \emph{permutation importance} \cite{breiman2001random}, where the values for a feature are permuted throughout the dataset and the effect on model performance is measured to indicate its importance.

\section{Methods}
\label{sec:methods}

\begin{figure}[!t]
    \centering
\includegraphics[width=\linewidth]{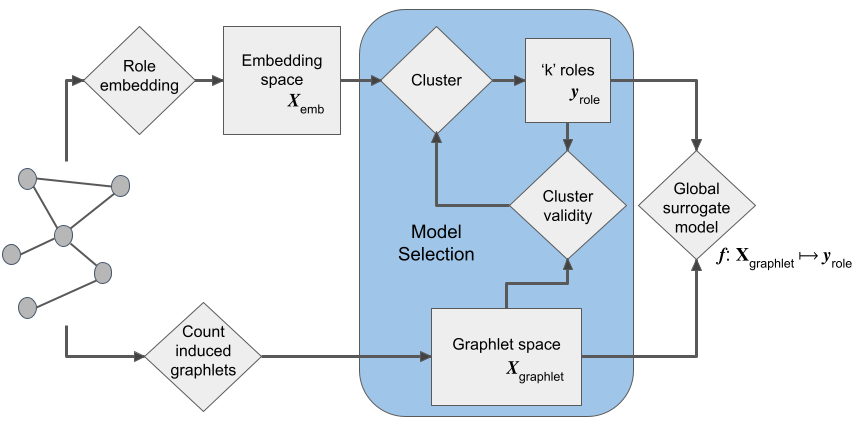}
    \caption{Overview of the workflow for the proposed framework.}
    \label{fig:workflow}
\end{figure}

In this section we outline a general framework for uncovering and evaluating structural roles on a graph. The complete process is illustrated in Figure \ref{fig:workflow}. 

We begin with a graph $G = (V,E)$, from which we wish to identify a set of discrete structural roles. We employ a role embedding algorithm to map each node $v \in V$ to an embedding space $X_{emb} = \mathbb{R}^{128}$, thus a clustering of $X_{emb}$ is considered a set of discovered roles $y_{role}$. Additionally, we represent the same nodes in the graphlet space $X_{graphlet} = \mathbb{Z}^{72}$ derived from $G$, where each node $v$ is represented by a \emph{bag-of-orbits} vector $x^v = \{x_0^v, x_1^v, ..., x_{72}^v\}$, with $x_i^v$ denoting the number of times node $v$ is involved in induced graphlet orbit $i$. We use a vocabulary of 72 orbits, which we count on graphlets of size 2 to 5 using the ORCA method \cite{hovcevar2017combinatorial}. We refer to graphlets and orbits according to the enumeration by \cite{prvzulj2007biological}. The graphlet space is first used to validate any set of roles we identify in the graph. By clustering the embedding space, we group nodes into $k$ discrete roles, which we can evaluate using cluster validity metrics calculated on the graphlet space. Employing various role embedding algorithms and clustering methods, we identify a set of candidate clusterings (or roles) according to the separation they achieve in the graphlet space. 

Graphlet-orbit counts are a powerful language for describing the structural features of nodes on a graph \cite{cunningham2013motifs,prvzulj2007biological}. As such, a clustering (according to node embeddings) which is valid in the graphlet-orbit space should offer a reasonable set of structural roles. Additionally, we employ graphlet orbits to interpret a set of discovered roles. For a candidate clustering $y_{role}$, we fit a surrogate model $f:X_{graphlet}\mapsto y_{role}$. By modelling the role assignment in the graphlet space, we can explore the feature importance and effect of the different graphlet orbits in role assignments, according to many \emph{model-agnostic} explanation techniques from the field of XAI \cite{molnar2020interpretable}. In Section \ref{sec:application}, we demonstrate graphlet-based explanation using permutation importance \cite{breiman2001random} and accumulated local effects plots \cite{apley2020visualizing}. As we will show, highlighting important or discriminatory orbits can offer a visual means of understanding the structure of a role in the graph.

\section{Application}
\label{sec:application}

In this section we apply the methods described in Section \ref{sec:methods} to extract and interpret sets of structural roles in a citation network. In the case of each paper, we compute Rao-Stirling diversity scores to indicate an articles interdisciplinarity and thus explore the distributions of IDR scores assigned to papers in different roles. We identify a set of roles which has grouped papers according to their interdisciplinarity. Finally, we use graphlets to interpret the structure of these more interdisciplinary roles, and highlight certain citation structures which are specific to interdisciplinary research. 

\subsection{Data}
In order to discover the citation structures of interdisciplinary research, we require a large, dense citation network that contains research from a diverse set of disciplines. In addition, we require that each paper can be assigned to a subject category or discipline, according to an established taxonomy. 
We construct a novel citation network using Microsoft Academic Graph \cite{sinha2015overview} citation data from a seed set of journal papers. This sets consist of samples of articles from Scopus indexed journals, stratified according to their All Science Journal Categories (ASJC). The graph contains samples of 1,500 articles published between 2017 and 2018 in Scopus indexed journals with the ASJCs \emph{`Computer Science', `Mathematics', `Medicine', `Chemistry', `Social Sciences', `Neuroscience', `Engineering'}, and \emph{`Biochemistry, Genetics and Molecular Biology'}. We maximise the completeness of the graph by including all available referenced articles that are published in Scopus indexed journals. In this manner, we produce a dense, multidisciplinary citation network, such that each article can be categorised according to the ASJC of the journal in which it was published. Later, these discipline categories can be used to compute article interdisciplinarity according to Rao-Stirling diversity of disciplines identified in both an articles citing and cited papers. In total, the citation graph contains 41,895 papers (nodes) and 129,159 citations (undirected edges). 

\subsection{Role Discovery}
For each article in the citation graph, we learn role embeddings according to 4 approaches: (i) \emph{Role2Vec} \cite{ahmed2019role2vec}, (ii) \emph{Struc2Vec} \cite{ribeiro2017struc2vec}, (iii) \emph{RolX} \cite{henderson2012rolx}, (iv) \emph{GraphWave} \cite{donnat2018learning}, and we cluster each embedding space using $k$-means clustering for values of $k$ between 2 and 19. Articles clustered according to their role embeddings represent a set of structural roles. 

\subsection{Role Interpretation}

\subsubsection{Validation.}
\begin{figure}[h!t]
    \centering
\includegraphics[width=0.8\linewidth]{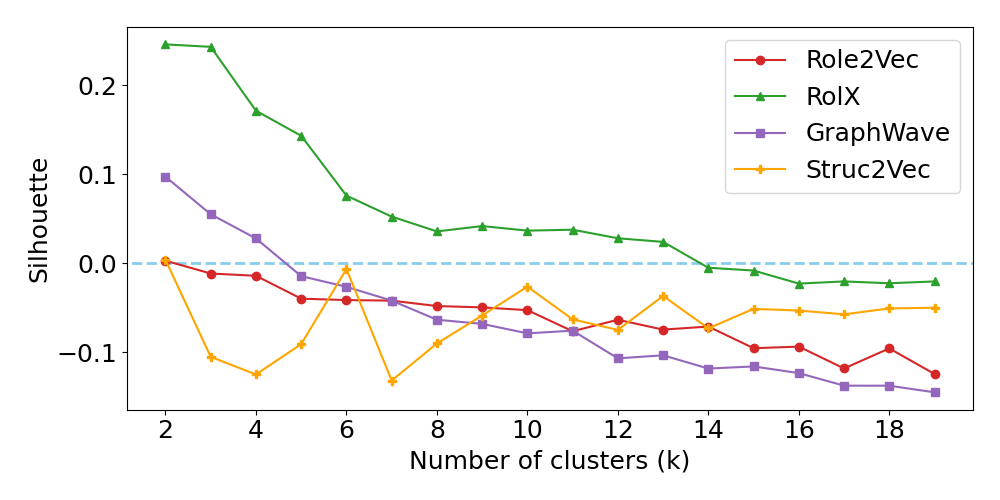}
\vskip -1em
    \caption{Silhouette score for $k$-means clusterings of the different embedding spaces. Silhouettes are calculated according the clusterings if we describe each node by their bag-of-orbits vector, rather than their embedding.}
    \label{fig:silhouettes}
\end{figure}

Figure \ref{fig:silhouettes} shows the cluster validity of the roles discovered according to the 4 role embedding methods. Each set of roles represents a $k$-means clustering of the embedding space, which is then transformed to the graphlet space, where we measure the validity of the clustering using silhouette score \cite{rousseeuwW1987silhouettes}. We plot the silhouette score for each embedding for values of $k$ in the range $[2,19]$. Silhouette scores can take a value in the range [-1,1]. A high score represents dense, separate clusters, a score of 0 indicates an overlapping clustering, while a negative score indicates an incorrect clustering. 
According to these scores, we choose 3 candidate roles to demonstrate interpretation and explanation: (i) Struc2Vec ($k = 6$) which is an outlier in the Struc2vec roles and achieves an overlapping clustering, (ii) RolX ($k = 3$) which has the highest silhouette score for all approaches with more than 2 clusters, and (iii) GraphWave ($k = 3$) which achieves a positive score.

\subsubsection{Surrogate models.}
\vskip-2em
We fit a random forest classifier to model role assignments according to the bag-of-orbits vectors for each node. That is, we learn a function $f:X_{graphlet}\mapsto y_{role}$ for each of the candidate roles $y_{role}$. We choose a random forest classifier as we anticipate that structural roles may be non-linear in the graphlet space and may rely on combinations and interactions of features to model higher-order structures. As the orbit counts follow a power-law distribution, we log-transform all features in the graphlet space. Table \ref{tab:permutation_importance} reports the 5 most informative features for each model according to permutation importance. In the case of the Struc2Vec ($k = 6$) roles, the overlap between clusters in the graphlet space is evident. The only informative features (with non-zero permutation importance) are small, local orbits -- the approach is blind to deeper, more complex structures. The RolX ($k = 3$) approach, which best separates the nodes in the graphlet space, has grouped nodes according to larger, simple structures. The most informative orbits: 15, 4, and 1, each refer to chains of varying length. Finally, GraphWave ($k = 3$) appears to have grouped the nodes according to more complex, higher-order structures. Many of the features important to role classification in the GraphWave case (27, 24, and 18) contain combinations of chains, stars and cliques. Depending on the domain or application in which we employ role discovery, any one of these sets of roles may be the most valid or useful. However, without modelling the role assignments in the graphlet space, we are unable to understand which structures are being grouped in the discovered roles. We will use the GraphWave roles to demonstrate further explanation and explore the structure of interdisciplinary papers.  

\begin{table}[]
    \centering
    \begin{tabular}{|>{\centering}p{0.5cm}|>{\raggedleft}p{3.5cm}|>{\raggedleft}p{3.5cm}|>{\raggedleft\arraybackslash}p{3.5cm}|}
\hline
{} & \centering Struc2Vec ($k = 6$) & \centering RolX ($k = 3$) & \multicolumn{1}{c|}{GraphWave ($k = 3$)} \\
\hline
1 &   0 (0.111 ±0.0005) &  15 (0.113 ±0.0016) &  27 (0.022 ±0.0006) \\
2 &   2 (0.010 ±0.0001) &   4 (0.007 ±0.0002) &  15 (0.014 ±0.0002)  \\
3 &   3 (0.006 ±0.0001) &   1 (0.006 ±0.0003) &  17 (0.013 ±0.0006)  \\
4 &   5 (0.002 ±0.0001) &  27 (0.005 ±0.0002) &  24 (0.013 ±0.0004) \\
5 &  16 (0.000 ±0.0000) &  19 (0.004 ±0.0003) &  18 (0.013 ±0.0004) \\
\hline
\end{tabular}
\vskip 0.5em
    \caption{Top 5 most important features for each surrogate model. Orbits are ranked by permutation importance, which is included in parentheses.}
    \label{tab:permutation_importance}
\end{table}

\subsubsection{Roles and their structure.}
\begin{figure}[!t]
    \centering
    \includegraphics[width=\linewidth]{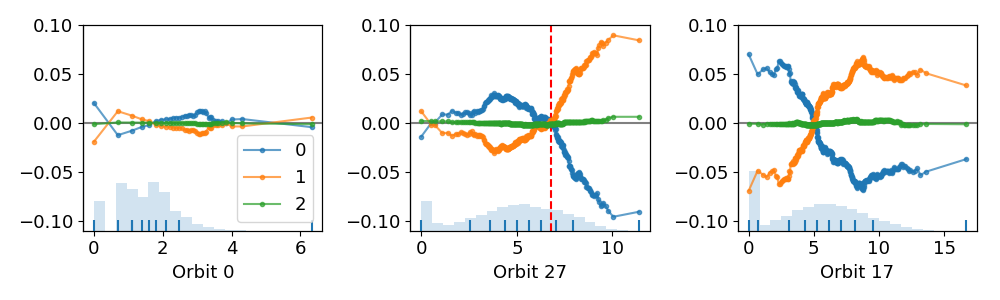}
    \vskip -0.3em
    \caption{Accumulated Local Effect \cite{apley2020visualizing} plots for a surrogate model which classifies nodes to GraphWave roles according to graphlet orbit counts. The figure shows the effect of 3 features: orbits 27 and 17; the most important features as measured by permutation importance, and orbit 0 (node degree); which we include for reference. In the case of orbit 27, we highlight the maximum value of orbit 3 that was observed in the graph.}
    \label{fig:ale}
\end{figure}

\begin{figure}[h!t]
    \centering
\includegraphics[width=\linewidth]{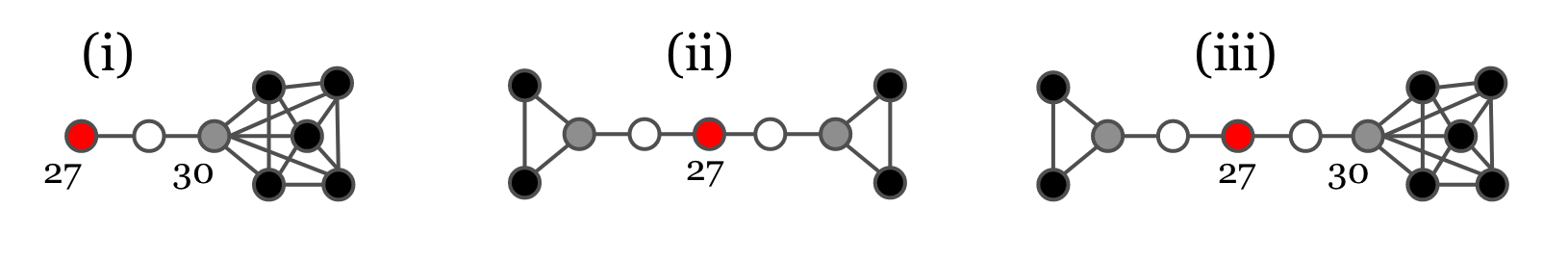}
    \caption{Higher order structures containing orbit 27. Here (i) and (ii) represent likely structures for nodes with high counts for orbit 27. When the count for orbit 27 exceeds a threshold, we infer structure (iii). }
    \label{fig:graphlet_explanation}
\end{figure}

\vskip-3.5em
The GraphWave method clusters 35,136 papers into role 0, 16,453 papers into role 1, and 306 papers into role 2. Figure \ref{fig:ale} contains Accumulated Local Effects (ALE) \cite{apley2020visualizing} plots for 3 features/orbits and their effect on classification to each of the 3 GraphWave roles. We illustrate the ALE of orbits 27 (the end of a chain adjacent to a clique) and 17 (the middle of a long chain) as two of the most important structures (according to permutation importance). We also include the ALE of orbit 0 (node degree) as a valuable reference; it is useful to confirm that the roles are indeed separating nodes according to more complex features and not simply the number of edges. The ALE plot for orbit 27 shows that for low-to-mid values of that orbit count, a node will be classified as role 0. However, if a node's count for orbit 27 exceeds a threshold, it will be classified as role 1. We suppose 2 scenarios when a focal node's count for orbit 27 (a chain adjacent to a clique) will become large: (i) the node is adjacent to a large community -- each triangle in which the node at position 30 participates will increase the count, (ii) the node exists at the center of a barbell graphlet, i.e. on a longer chain between two or more communities communities. We illustrate these scenarios in Figure \ref{fig:graphlet_explanation}. There should exist some threshold value of 27 beyond which a node must exist on the chain between two communities. For example, if a focal node has a count for orbit 27 that is greater than the count of triangles (orbit 3) for the node at position 30, then the focal node must be adjacent to a second community (scenario (iii)). This threshold will be represented by the greatest value of orbit 3 in the graph. We include this value for reference in Figure \ref{fig:ale}. Beyond this threshold, a node is more likely to be classified in role 1. Accordingly, we conclude that a node/paper that is on the end of a chain adjacent to a community will be assigned to role 0, while a node that exists on a bridge between two communities will be assigned to role 1. 

\begin{figure}[!t]
    \centering
    \includegraphics[width=\linewidth]{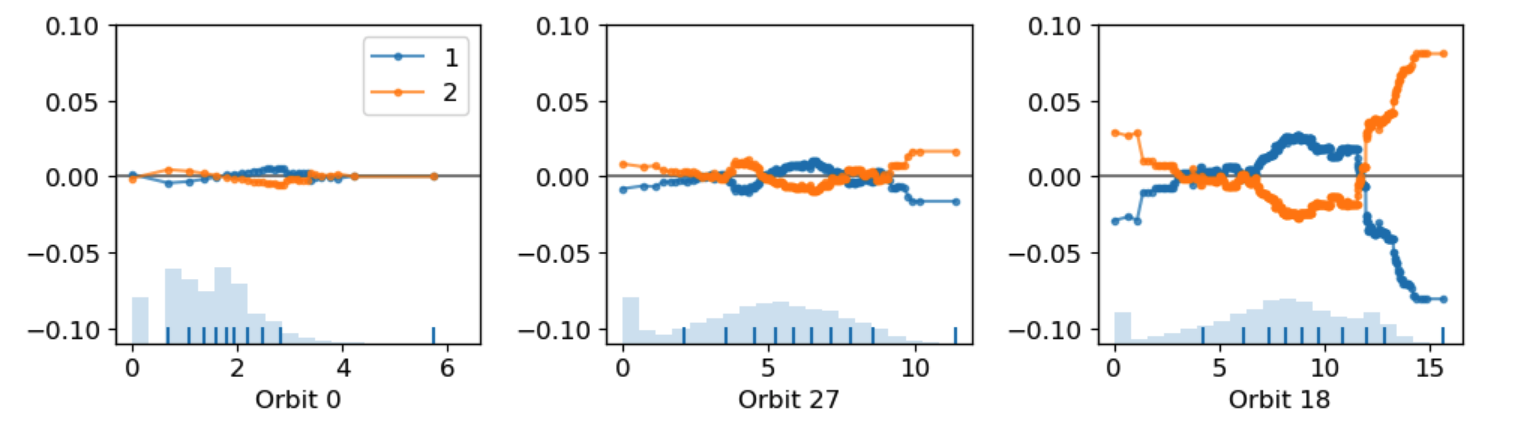}
    \vskip -0.3em
    \caption{Accumulated Local Effect \cite{apley2020visualizing} plots for a surrogate model which classifies nodes to GraphWave roles 1 or 2 according to graphlet orbit counts. The figure plots the effect of 3 features: orbits 0; node degree and 27; previously the most important feature for the global model, and orbit 18; the most important feature in the surrogate model.}
    \label{fig:ale_2}
\end{figure}

In order to identify the structures specific to the smallest role (role 3), we can fit another surrogate model on only the nodes in cluster 1 and 2. ALE plots for this model are included in Figure \ref{fig:ale_2}. In this case we find that orbit 27 does not meaningfully distinguish between the two roles. Instead, orbit 18 (the end of a chain adjacent to a star) is the most informative feature, and, for very high values of this orbit count, nodes will be assigned to role 2. Such nodes likely represent the centre of a bridge between large communities that are less densely connected (i.e., containing many open triads). We conclude this to be an important structure for role 3. 

\subsection{Interdisciplinary Roles}

\begin{figure}[!t]
    \centering
\includegraphics[width=\linewidth]{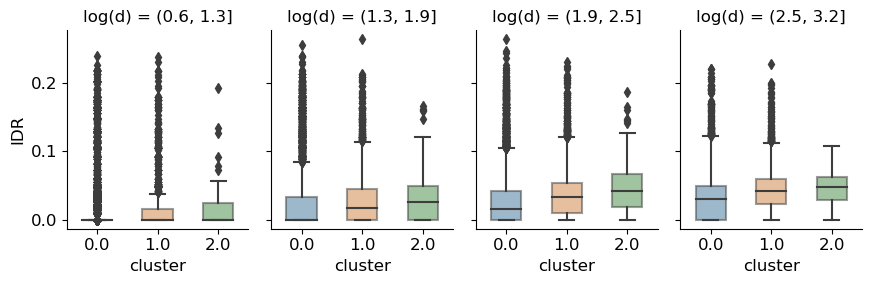}
    \caption{Interdisciplinarity scores (IDR) computed for papers in each of the 3 GraphWave clusters. IDR is computed as the Rao-Stirling \cite{stirling2007general} diversity of the research disciplines identified in an articles citing papers. We bin papers/nodes according to the log of their degree, and compare distributions within each bin.}
    \label{fig:idr}
\end{figure}

Figure \ref{fig:idr} plots the IDR scores for the papers in each of the GraphWave roles. IDR is calculated according to the Rao-Stirling diversity of the ASJC categories identified in an article's citing papers. As Rao-Stirling IDR scores may be biased according to the number of articles in the summation, we bin nodes by degree, and plot IDR distributions for each role, within each bin. Specifically, we log-transform the node degrees and group nodes into 10 bins of equal width, within which we plot IDR distributions for each role if the bin contains more than 50 papers from each role. According to these plots, we note that the structural roles identified by GraphWave have grouped papers into clusters with different IDR distributions. Even when we account for node degree (a potential bias of the Rao-Stirling IDR score), papers assigned to structural roles 1 and 2 have consistently greater IDR distributions than those assigned to role 0. We recall some of the important structures that were identified for roles 1 and 2: (1) a bridge between densely connected communities, and (2) a bridge between large, sparsely connected communities. We conclude these to be citation important structures associated with interdisciplinary research.  

\section{Discussion and Conclusions}

Many modern methods for role discovery in graphs rely on node embeddings \cite{rossi2020proximity}. We have applied a number of these methods to learn structural roles on a research citation network. While these methods have previously been shown to be capable of grouping nodes into known roles (e.g., in synthetic graphs or transport networks \cite{ribeiro2017struc2vec,donnat2018learning}), it remains unclear how roles could be understood or validated when applied to graphs with unknown roles. In this work we outlined a framework for interpreting a set of discovered roles using graphlets and orbits. We leveraged methods from the field of explainable AI to explore the subgraph structures that are specific to discovered roles. We demonstrated our approach on a citation network, where we identified important structures specific to interdisciplinary research. It is evident in our analysis that different role discovery methods learn different sets of structural roles. In different applications many or all of these clusterings may be valid, but it is critical that we can compare the roles discovered by different methods. While our framework is general, and applicable to explanation and validation in all role discovery tasks, we highlight the utility of structural role embeddings in mapping interdisciplinary research. 

For the task of identifying and mapping IDR interactions, structural paper embeddings could be augmented by considering additional, non-structural information, such as article or abstract text. This could provide a richer paper representation, without imposing a predefined or static disciplinary classification on the graph. There is also scope for improving upon our proposed framework. Many \emph{model-agnostic} approaches have been developed for explaining surrogate models \cite{molnar2020interpretable}, which could be applied to interpret the role assignments in the graphlet space. For example, second-order effects of pairs of features can be calculated in a similar manner to the ALE analysis we have included \cite{apley2020visualizing}. Combinations of graphlets could be highly effective in modelling higher-order, more complex graph structures. One possible limitation of our current framework is the number of correlated features in the graphlet space. In future applications it may be possible to reduce the set of graphlet orbits to a minimal set of uncorrelated features via traditional feature selection techniques. 


%
%
\bibliographystyle{abbrv}
\bibliography{references}

\end{document}